\documentstyle[pre,aps,epsf,twocolumn]{revtex}
\begin{document}
\draft
\title{Phase diagram of a hard-sphere system in a quenched random potential:
a numerical study}
\author{Chandan Dasgupta\cite{chandan}}
\address {Centre for Condensed Matter Theory, 
Department of Physics, Indian Institute of Science, Bangalore 560012
India}
\author {Oriol T. Valls\cite{oriol}}
\address{School of Physics and Astronomy and Minnesota Supercomputer Institute,
University of Minnesota,
Minneapolis, Minnesota 55455-0149}
\date{\today}
\maketitle
\begin{abstract}
We report numerical results for the phase diagram in the
density-disorder plane of a hard sphere system in the presence of
quenched, random, pinning disorder. Local minima of a discretized
version of the Ramakrishnan-Yussouff free energy functional are located
numerically and their relative stability is studied as a function of
the density and the strength of disorder. Regions in the phase diagram
corresponding to liquid, glassy and nearly crystalline states are
mapped out, and the nature of the transitions is determined. The liquid
to glass transition changes from first to second order as the strength
of the disorder is increased. For weak disorder, the system undergoes a
first order crystallization transition as the density is increased.
Beyond a critical value of the disorder strength, this transition is
replaced by a continuous glass transition.  Our numerical results are
compared with those of analytical work on the same system.
Implications of our results for the field-temperature phase diagram of
type-II superconductors are discussed.

\end{abstract}
\pacs{64.70.Pf, 64.60.Ak, 64.60.Cn}

\section{Introduction}
The equilibrium phase diagram of a classical system of interacting
particles in a quenched, random, pinning potential is an important
subject on which much effort is currently being spent~\cite{rev}.
There are several experimentally studied systems, such as vortices in
the mixed phase of high-T$_c$ superconductors~\cite{htsc}, fluids
confined in porous media~\cite{pormed}, magnetic bubble
arrays~\cite{magbub}, and Wigner crystals~\cite{wxtal}, which provide
physical realizations of a collection of interacting classical objects
in the presence of an external, time-independent, random potential. In
the absence of such a potential, systems of this kind are expected to
crystallize at sufficiently low temperatures. Several years ago,
Larkin~\cite{larkin} showed that the presence of arbitrarily small
amount of random pinning disorder destroys long-range translational
order in all physical dimensions $d < 4$.  However, recent theoretical
studies~\cite{Nat,GiaLeD} suggest that weak disorder distorts the
crystalline state only slightly, leading to a phase with perfect
topological order and logarithmic fluctuations of the relevant
displacement field. This phase, with quasi-long-range translational
order and power-law Bragg peaks in the structure factor, is called a
``Bragg glass''~\cite{GiaLeD}. The transition point between a Bragg
glass and the high-temperature liquid phase is likely to be shifted
with increasing disorder, but the transition is believed to remain
first order as long as the disorder is weak.  A question of obvious
interest is how this transition temperature and the nature of the
transition depend on the strength of the random potential.

As the relative strength of the disorder is increased, so that the
week-disorder situation described above no longer applies, the Bragg
glass phase is expected to undergo a transition to a topologically
disordered amorphous phase with only short-range translational
correlations. It is not yet clear whether this phase is
thermodynamically distinct from the high-temperature liquid. An
interesting possibility is that it is analogous to the glassy phase
obtained by supercooling a liquid below the structural glass transition
temperature in the absence of external quenched disorder~\cite{caveat}.
If this is so, then the phase diagram of such systems would contain
three phases: a Bragg glass phase obtained at low temperature and weak
disorder, an amorphous (without quasi-long-range translational order)
glassy phase at low temperatures and strong disorder, and a weakly
inhomogeneous (because
of the presence of the external random potential)
liquid phase at high temperatures. The glassy phase would be a
thermodynamically stable one in these systems. This is different from
the situation in the absence of external disorder where the crystalline
state is the true equilibrium state near the structural glass
transition and both the supercooled liquid and the glass are
metastable. In other words, the presence of external disorder may lead
to the possibility of occurrence of a true, thermodynamically stable,
glassy phase.

The phase diagram~\cite{htsc} in the temperature ($T$) -- magnetic
field ($H$) plane of layered, highly anisotropic, type-II
superconductors such as ${\rm Bi_2Sr_2CaCu_2O_8}$ in a magnetic field
perpendicular to the layers is a credible candidate to exhibit these
three phases.  For a wide range of values of $H$, the flux lines in
these materials may be regarded as columns of interacting ``pancake''
vortices~\cite{htsc} residing on the layers, and the properties of the
mixed phase may be described in terms of the classical statistical
mechanics of these point-like objects. In these compounds, at low
enough fields, a flux-lattice melting transition separates a nearly
crystalline state of the flux lines from a disordered ``vortex liquid''
state. The first-order character of this transition has been carefully
documented~\cite{Zeldov}.  When the magnetic field is increased, the
transition becomes continuous~\cite{Zeldov,Safar}, and the nearly
crystalline state appears to be replaced by an amorphous state called
``vortex glass''~\cite{F2-Huse} which is endowed with glassy properties
such as non-ohmic current-voltage characteristics~\cite{Koch}.  It is
generally assumed~\cite{F2-Huse} that the vortex glass phase owes its
existence to the presence of point-like pinning disorder.  Observation
of Bragg peaks in neutron scattering experiments~\cite{Cubitt} confirms
that the phase at low $H$ and $T$ is a Bragg glass.  As the effective
strength of the disorder is increased, either indirectly by increasing
$H$ (which is believed to increase~\cite{Safar} the
effective strength of the disorder), or directly by increasing the amount
of defects in the sample~\cite{Khaykovitch}, the Bragg glass phase
changes over to the vortex glass. The latter is separated from
the liquid by a continuous transition~\cite{comment}.  This phase
diagram, thus, suggests that the first-order liquid-to-crystal
transition in a three-dimensional (3d) system of point-like objects may
be driven by the pinning disorder into a continuous liquid-to-glass
transition.

The formation of a glassy phase at strong disorder was investigated
recently~\cite{thal} in an analytic study of the phase diagram of a
system of hard spheres in a random pinning potential of arbitrary
strength. This work used a combination of two ``mean-field''- type
approaches based on the ``replicated liquid
formalism''~\cite{pormed,MenDas,MezPar}: the replica method~\cite{book}
was used for treating the effects of quenched disorder, and the
hypernetted chain approximation~\cite{hm86} 
to calculate the equilibrium correlation
functions in the liquid in the presence of the pinning potential. These
correlation functions were then the input in a replicated density
functional~\cite{MenDas} of the Ramakrishnan-Yussouff (RY)
form~\cite{ry} from which the location of the freezing transition of the
liquid into a nearly crystalline (Bragg glass) phase was obtained. 
The possibility
of a liquid-to-glass transition was investigated using
the phenomenological approach of M\'ezard and
Parisi~\cite{MezPar}.
The resulting\cite{thal}
phase diagram in the density -- disorder plane (the density, rather
than the temperature, is the appropriate control parameter for a
hard-sphere system)  shows three
thermodynamic phases: a nearly crystalline Bragg glass, an
amorphous glassy phase, and a low-density liquid. It is
consistent with the expectation (from earlier
work~\cite{MenDas} and the Lindemann criterion~\cite{ErtNel})
that the density at which the Bragg glass to
liquid transition occurs should move to higher values as the strength
of the disorder is increased.   
The first-order crystallization transition is replaced by
a continuous glass transition as the disorder strength is increased
above a threshold value. This phase diagram 
is thus qualitatively similar to that proposed for some
layered type-II superconductors if, as noted above,
the density is replaced by the
temperature $T$  and the disorder strength 
by the magnetic field $H$.

In the present paper, we report the results of a numerical investigation of
the phase diagram of the same system, i.e. a hard-sphere fluid in the
presence of a random pinning potential with short-range spatial
correlations. Our work involves the use of direct numerical
minimization to study the effects of the presence of an external random
potential on the minima of a discretized version of the RY free energy
functional for the hard-sphere system. It is known~\cite{cd92} that in
the absence of external disorder, this model free energy functional
exhibits, at sufficienly high densities, a large number of ``glassy''
local minima characterized by inhomogeneous but aperiodic density
distributions. In addition, a global minimum corresponding to the
crystalline solid is also found at high densities if the sample size
and the discretization scale are commensurate with the crystal
structure. If they are incommensurate, only the glassy minima are
present. We have carried out extensive numerical investigations of the 
resulting free-energy landscape~\cite{cdotv96,cdotv98,land,sit} 
in the absence of disorder.
In the present study, and with the physical situations described above
in mind, we develop similar numerical methods to find the location and
structure of the local minima of the same model free energy with the
addition of the presence of a time-independent, random, one-body
potential.

Using these numerical methods we investigate how the uniform liquid,
crystalline solid and glassy minima of the free energy in the absence
of the random potential evolve as the strength of the random potential
is increased from zero. We also examine the dependence of the free
energy of these minima, and of the density structure (as given by the
two-point density correlation function) of the system at these minima,
on the strength of the disorder. In this picture, a transition from one
phase to another is signalled by the crossing of the free energies of
the corresponding minima of the free energy. By monitoring where these
crossings occur as the density and the strength of the disorder are varied,
we are able to map out the phase diagram in the density -- disorder
plane. This phase diagram is qualitatively very similar to the one
obtained in the analytic work. For weak disorder we
find, in the commensurate case as described above where a crystalline
minimum exists, a first-order liquid-to-crystal transition that moves
to higher density as the strength of the disorder is increased. In the
metastable ``supercompressed'' regime (i.e. at a density higher than
the value at which equilibrium crystallization takes place for the
commensurate case), we find in all cases a liquid-to-glass transition.
The density at which this transition occurs decreases (very slowly for
the largest systems studied, which are incommensurate, and more rapidly
for the smaller, commensurate systems) as the strength of the disorder
is increased. The nature of this glass transition depends on the strength
of the disorder: it is first order when the disorder is weak, but it changes
to second order beyond a certain critical value of the disorder
strength.  For the commensurate case, the crystallization line crosses
the glass transition line at or very near the same critical value of
the disorder strength, so that the system at stronger disorder then
undergoes a liquid-to-glass transition (instead of the
liquid-to-crystal transition found for weak disorder) as the density is
increased from a low value. The continuous nature of the glass
transition in the large disorder regime is in contrast with the first
order transition from the liquid to a crystalline or glassy state
(depending on the commensurability) at small values of the disorder
strength. Thus, this work provides support to the prediction that the
first-order liquid-to-crystal (Bragg glass) transition should change
over to a continuous liquid-to-glass transition as the strength of the
pinning disorder is increased above a critical value.

The rest of the paper is organized as follows. In section II, we define
the model studied here and outline the numerical procedure used in this
study. The numerical results obtained for the different transition
lines in the density--disorder plane are described in detail in section
III. Section IV contains a summary of our main results and a few
concluding remarks.

\section{Methods}
\label{methods}
\subsection{The Free Energy functional}

As discussed in the Introduction, our starting point is the free energy
as a functional of the time-averaged local density $\rho({\bf r})$ at each
point ${\bf r}$. We write this free energy functional in the form:
\begin{equation}
F[\rho]=F_{RY}[\rho]+F_s[\rho]
\label{fe}
\end{equation}
where the first term 
in the right-hand side is the RY free enrgy
functional~\cite{ry} for hard spheres in the absence of disorder, and the
second is the contribution
arising from the presence of a quenched random potential. Thus
we have:
\begin{eqnarray}
\beta F_{RY}[\rho] &=& \int{d {\bf r}\{\rho({\bf r})
\ln (\rho({\bf r})/\rho_0)-\delta\rho({\bf r})\} }  \nonumber \\
&-&\frac{1}{2}\int{d {\bf r} \int {d{\bf r}^\prime
C({|\bf r}-{\bf r^\prime|}) \delta \rho ({\bf r}) \delta
\rho({\bf r}^\prime)}} .
\label{ryfe}
\end{eqnarray}
Here, we have defined $\delta \rho ({\bf r})\equiv \rho({\bf r})-\rho_0$ as the
deviation of  ${\rho(\bf r})$ from $\rho_0$,
the density of the uniform liquid, and 
taken the zero of the free energy at its uniform liquid value. 
In Eq.(\ref{ryfe}), $\beta=1/(k_B T)$, 
$T$ is the temperature and the function $C(r)$ 
is the direct pair correlation
function~\cite{hm86} of the uniform liquid at density $\rho_0$, which
can be analytically expressed in terms of the usual dimensionless density 
for hard spheres of diameter $\sigma$,
$n^*\equiv \rho_0 \sigma^3$, by making use of the Percus-Yevick
approximation~\cite{hm86}. This approximation  is
sufficiently accurate in the density ranges ($n^* \le 1.0$) considered
in this paper. We write also:
\begin{equation}
\beta F_s[\rho]= \int{d {\bf r} \delta \rho({\bf r}) V_s({\bf r})}
\label{sfe}
\end{equation}
where $V_s({\bf r})$ is an external  potential (in dimensionless form)
representing the random, quenched disorder.
We will assume that $V_s$ has zero mean and short-range Gaussian
correlations as detailed below. 

In order to carry out numerical work, we discretize our system. We
introduce for this purpose a simple cubic computational mesh of size
$L^3$ with periodic boundary conditions.
On the sites of this mesh, we define density variables 
$\rho_i \equiv \rho({\bf r}_i) h^3$, where $\rho({\bf r}_i)$ is the
density at site $i$ and $h$ the spacing of the
computational mesh. It is known from previous 
work~\cite{cd92,cdotv96} that in the absence of any random potential,
this discretized system crystallizes at sufficiently high
densities if the quantities $h$ and $L$ are
commensurate with a fcc structure with appropriate lattice spacing,
whereas no crystalline state exists when the computational mesh
is incommensurate with a fcc structure. Both commensurate and
incommensurate systems exhibit~\cite{cd92,cdotv96,cdotv98,land} 
many glassy (inhomogeneous but aperiodic)
minima of the free energy at densities higher than the value at which
crystallization occurs in commensurate samples. 

To model the random potential $V_s({\bf r})$, we introduce random
variables $\{V_i\}$ defined at the sites of the computational mesh. 
These variables are
uncorrelated with one another, and distributed according to a Gaussian 
probability distribution with zero mean and variance $s$. Thus, $s$ 
represents the dimensionless strength of the disorder. In terms of
these quantities, the dimensionless free energy of our discretized
system has the form
\begin{eqnarray}
\beta F &=& \sum_i \{\rho_i \ln (\rho_i/\rho_\ell) -
(\rho_i-\rho_\ell)\} \nonumber \\
&-&\frac{1}{2} \sum_i \sum_j C_{ij}(\rho_i-\rho_\ell)(\rho_j-\rho_\ell)
+\sum_i V_i (\rho_i-\rho_\ell),
\label{discr}
\end{eqnarray}
where the sums are over all the sites of the computational mesh,
$\rho_\ell \equiv \rho_0 h^3$, and $C_{ij}$ is the discretized form of 
the direct pair correlation function $C(r)$ of the uniform liquid.

Our objective is to study the phase diagram of this system in the
$(n^*,s)$ plane, in which in principle crystalline, liquid, and glassy
phases may be found. The thermodynamics of hard spheres
in the clean limit is determined by the dimensionless density $n^*$
only.  Our rescaling of the potential $V_s$ by $\beta$ (see
Eq.(\ref{sfe})) ensures that $s$ is now the only additional relevant
variable. We wish to locate various transition lines in this $(n^*,s)$
plane, that is, we wish to determine
which phase (crystal, glass or liquid)
is the thermodynamically stable one at different points in this plane.
We also wish to know when and how a certain phase becomes metastable or
unstable as we move around in the $(n^*,s)$ plane. In our mean-field
description, different phases are represented by different minima of
the free energy. If several local minima of the free energy are
simultaneously present, then the minimum with the lowest free energy
represents the thermodynamically stable phase and the other local
minima correspond to metastable phases. A crossing of the free energies
of two different minima
signals a first-order phase transition. 
The point where a minimum becomes locally unstable (i.e. changes from
a true minimum to a saddle point or disappears altogether) 
corresponds to a
mean-field spinodal point representing the limit of metastability of
the corresponding phase. A merging of the transition point with the
spinodal points of the two phases signals a continuous phase transition
in this description. Thus, a study of how the minima of the free energy of
Eq.(\ref{discr}) evolve as $n^*$ and $s$ are changed is sufficient for 
mapping out the mean-field phase
diagram in the $(n^*,s)$ plane.

We locate the minima of the free energy by using
a numerical procedure generalized from
that originally developed for the clean case~\cite{cd92}. This procedure
works by changing the local density variables $\{\rho_i\}$ in a
way that ensures that these changes always decrease the free energy. 
Given an initial configuration of the variables $\{\rho_i\}$, this
procedure finds, by constantly moving downhill on the free-energy
surface in the multidimensional configuration space spanned by the
$L^3$ variables $\{\rho_i\}$, the local minimum whose 
basin of attraction contains the
intial state. Thus, different local minima of the free energy can be 
located by using this minimization procedure for different,
appropriately chosen, initial configurations. 

As noted earlier, there are in our system
three different kinds of free-energy
minima: liquid, crystalline and glassy. In the clean limit ($s=0$), it
is easy to distinguish among them: the liquid
minimum has uniform density ($\rho_i=\rho_\ell$ for all $i$), the
crystalline minimum has a periodic distribution of the density
variables ($\rho_i$ is close to unity at mesh points corresponding to
the sites of a fcc lattice, and close to zero at all other mesh
points), and a glassy minimum exhibits a strongly inhomogeneous 
nonperiodic density
distribution (some of the $\rho_i$'s are close to unity and the others
are close to zero). This symmetry-based
distinction among minima of different kind becomes less clear when the
external random potential is turned on: for $s \ne 0$, the density
distribution in the liquid phase is not completely homogeneous, and the
crystalline state is not strictly periodic. 

We use here, therefore,
a procedure of ``adiabatic continuation'' to distinguish
among the liquid, crystalline and glassy minima in the presence of the
disorder. This 
procedure works as follows: We start with a minimum of a particular
kind obtained at $s=0$ for a given value of $n^*$. 
There is no difficulty 
in generating the liquid (and if appropriate the
crystalline) configuration for the pure system. Glassy states at $s=0$
are easily
obtainable also, in the right density ranges, by the procedures
described in Ref. \onlinecite{land}. Indeed, we have used
in many cases the same density configurations obtained there,
which were available as computer files. After thus
choosing the initial state,
we generate a set of uncorrelated random numbers
$r_i,\,i=1,\ldots,L^3$, 
distributed according to a Gaussian with unit variance. A
``realization'' of the random potential $\{V_i\}$ is obtained by
multiplying these random numbers by the strength parameter $s$. The
initial $s=0$ minimum is then ``followed'' to finite $s$ by increasing
$s$ in small steps  $\delta s$ ~\cite{nosmall}. 
After each step increase,
the minimization routine is run, to find the nearest local minimum.
Thus, in
the first step of this process, the initial configuration is that of
the minimum obtained at $s=0$ and the values of $V_i$ are set at
$(\delta s) r_i$. The resulting density configuration obtained from the
minimization routine is then used as the initial configuration for the
next step, with the values of $V_i$ incremented to $2 (\delta s) r_i$. 
During this process, the random variables $\{r_i\}$ are held fixed --
only the strength parameter $s$ in increased in steps of $\delta s$. By
iterating this procedure, minima of different kinds obtained at $s=0$ for
a certain $n^*$
are ``followed'' at
constant density to the desired value of $s$. We use the terms
``liquid'', ``crystalline'' and ``glassy'' to denote 
the continued $s \ne 0$ minima
obtained from a $s=0$ minimum of the corresponding kind by
using this continuation procedure without crossing transition lines.
We will see below that even at large $s$, there are distinguishable
differences in the structure of the different kinds of minima.

Once a minimum of the desired kind is obtained at a particular point in
the $(n^*,s)$ plane, the free energy at the minimum reached, as well as
the entire density configuration of the system at the minimum
are obtained and can be analyzed.  The 
translational correlations can
be quantified by the two-point correlation
function $g(r)$ of the density variables $\{\rho_i\}$. This function is
defined as
\begin{equation}
g(r) = \sum_{i>j} \rho_i \rho_j f_{ij}(r)/[{\bar{\rho}}^2
\sum_{i>j}f_{ij}(r)], 
\label{gofr}
\end{equation}
where the distance $r$ is measured in units of $\sigma$, 
$\bar{\rho} \equiv \sum_i \rho_i/L^3$ is the average value of the
$\rho_i$ variables 
at the minimum under
consideration, and $f_{ij}(r) = 1$ if the separation between
mesh points $i$ and $j$ lies between $r$ and $r+\Delta r$ ($\Delta r$
is a suitably chosen bin size),
and $f_{ij}(r) = 0$ otherwise. This function represents the spatial
correlation of the {\it time-averaged} local density, and is distinct
from the {\it equal-time}, two-point density correlation function which
is often called $g(r)$ in the literature.
We also calculate 
$\rho_{max}$, the maximum value of the $\rho_i$
variables at the minimum, which gives additional
information about the inhomogeneity when
contrasted with $\bar{\rho}$ or its rescaled
equivalent $\rho_{av}
\equiv \bar{\rho}(\sigma/h)^3$ at the minimum. 

In addition to examining the transitions by looking at
discontinuities in $F$, $g(r)$ and the density
configurations, we also  directly
check on the stability of the corresponding minima. The stability of a
local minimum requires that all the eigenvalues of the Hessian matrix
$\bf M$ whose elements are given by
\begin{equation}
M_{ij} \equiv \frac{\partial^2 (\beta F)}{\partial \rho_i \partial
\rho_j} = \frac{1}{\rho_i} \delta_{ij} - C_{ij}
\label{hess1}
\end{equation}
evaluated at the minimum must be positive. This matrix is difficult to
handle numerically if the minimum under consideration is strongly
inhomogeneous, with some of the $\rho_i$'s very close to zero. In such
cases, the $1/\rho_i$ in the first term on the right-hand side of
Eq.(\ref{hess1}) causes numerical difficulties. To avoid this problem, we
consider instead the closely related matrix ${\bf M}^\prime$ whose
elements are given by
\begin{equation}
M_{ij}^\prime \equiv \sqrt{\rho_i} M_{ij} \sqrt{\rho_j} = \delta_{ij} 
- C_{ij}\sqrt{\rho_i \rho_j},
\label{hess2}
\end{equation}
evaluated at the minimum under consideration. It is easy to show that 
an instability of the minimum corresponds to the vanishing of the
smallest eigenvalue $\lambda$ of this matrix. In our numerical work, we
calculate the value of $\lambda$ in order to check whether the minimum
under study becomes unstable as $n^*$ or $s$ is varied.

In our computations we have included the density range from $n^*=0.65$ to 
$n^*=0.95$, and values of $s$ from zero to about two. These are
sufficient to encompass the phenomena that we wish to study. We have
used three lattice sizes, $L = 12, 15$, and 25. For the last two we
have used an incommensurate ratio $h/\sigma=1/4.6$, whereas for the smallest
lattice we have taken the commensurate value $h/\sigma=0.25$.

\section{Results}
\label{res}

\subsection{General considerations: Phase diagram}

At a general point in the $(n^*,s)$ plane, the system may have a number
of local minima, one of which is the absolute minimum while the others
are metastable. Possible minima are that corresponding to the liquid
(the one with uniform density at $s=0$ and its continuation to finite
$s$), the crystalline minimum, by which we similarly mean the one with
a periodic structure at $s=0$, and its continuation as described in the
preceding section, and a large number of glassy minima. As the values of
$n^*$ or $s$ change, free energy minima may in general appear and
disappear, and the free energy values of those that remain  change.
There will therefore be a number of instabilities and transitions,
which are the main subject of our study.

Consider first the previously studied~\cite{cd92,land,sit} case of the
disorder-free system ($s=0$ line). There, only the uniform liquid
minimum is present at low densities. As $n^*$ increases, a crystalline
minimum appears if the computational mesh is commensurate. When $n^*$
is further increased, a density is reached at which the crystal becomes
thermodynamically stable, that is, its free energy becomes lower than
that of the liquid state. We will denote this density as $n^*_D$.
Regardless of commensurability, many glassy minima appear as the density is
further increased. We denote by $n^*_C$ the density at which the
first glassy minimum makes its appearance. Alternatively, one may
consider the evolution of the glassy minima as $n^*$ is {\it decreased} from
a large initial value, and define $n^*_C$ as the density at which the
last remaining glassy minimum becomes locally unstable and disappears: 
the free energy of
this last remaining glassy minimum crosses that of the liquid at a
density $n^*_B$ which is somewhat higher than $n^*_C$. This density
corresponds to a liquid to glass transition. In the commensurate case,
the density $n^*_C$ is above the crystallization density $n^*_D$, and
the free energy of the crystalline minimum is lower than that of the 
glassy minima. Thus, the glass transition in the pure system occurs in a
``supercompressed'' regime where the crystalline state is the
thermodynamically stable one. 

When we include the effects of the disorder ($s>0$), we find
yet another density, $n^*_A$, at which the liquid minimum becomes locally
unstable (i.e. ceases to exist as a local minimum of the free energy). 
For weak disorder, the value of $n^*_A$ is large (substantially higher than
$n^*_B$)  so that the four densities $n^*_A$,
$n^*_B$, $n^*_C$, and $n^*_D$ are in decreasing order. Thus, we  have
four (three in the incommensurate case where the crystalline state is
absent) functions $n^*_X(s)$ with $X=A,B,C,D$, representing precisely
the four transitions or instabilities defined above.  We denote the
corresponding lines in the $(n^*,s)$ plane as the $A,B,C,D$ lines.  The
determination of the location of these lines is one the main results of
our work. These results will be discussed below, but to fix ideas and
to make this discussion easier to follow, we show in Fig.~\ref{nfig1}
these four lines for the $L=12$ commensurate case. There,
\vbox{
\vspace{0.5cm}
\epsfxsize=8.0cm
\epsfysize=8.0cm
\epsffile{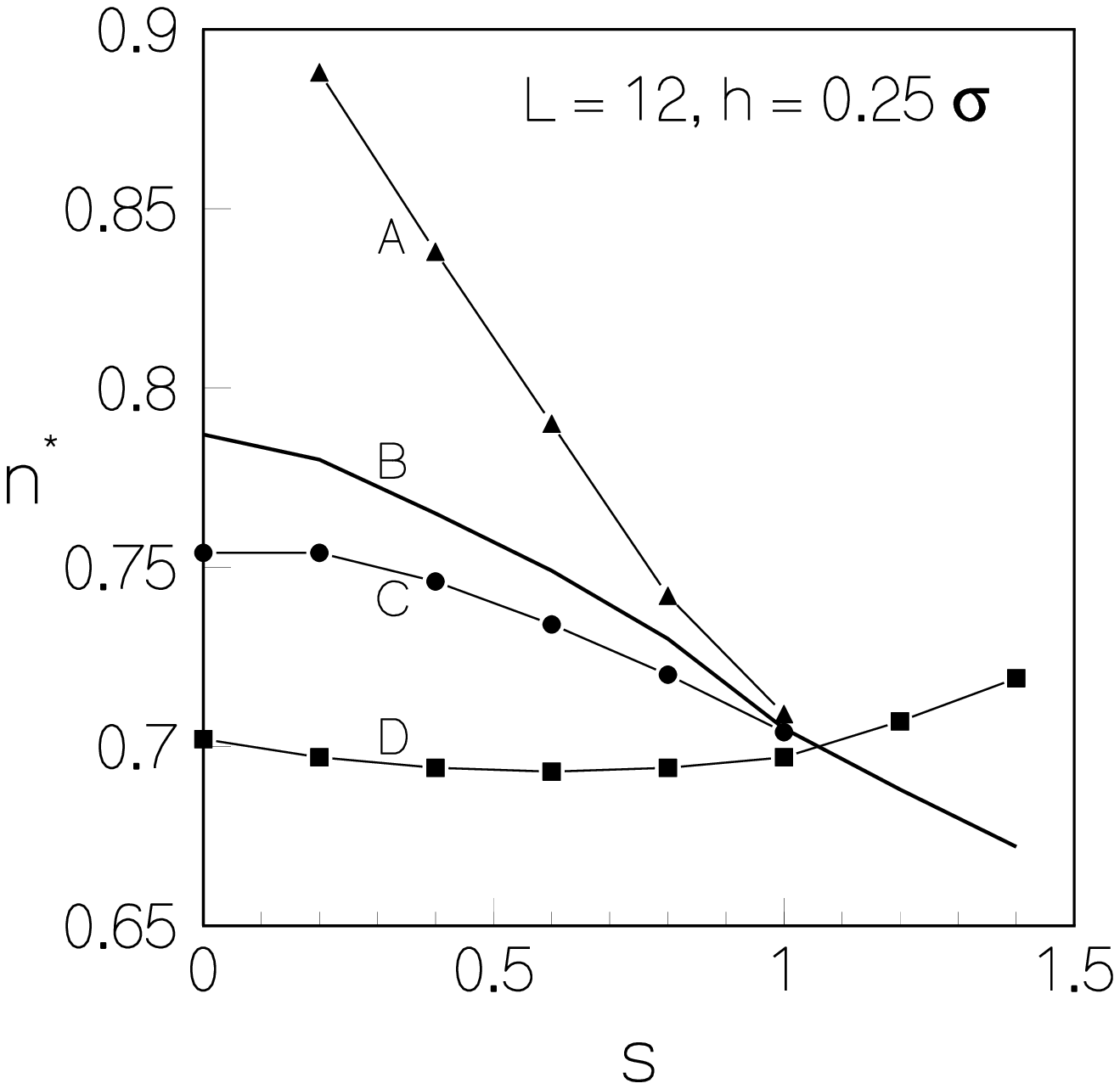}
\begin{figure}
\caption{The overall phase diagram of the hard sphere system in the
density ($n^*$) -- disorder ($s$) plane, obtained
for the $L=12$ commensurate sample. The meaning of the line labels is
explained in the text. The results shown are averages over 5
realizations of the disorder.} 
\label{nfig1}
\end{figure}}
the general structure of the phase diagram, including the
general shape of the four lines $n^*_X(s)$ can be seen. Similarly,
we show in Fig.~\ref{nfig1b} the three lines $n^*_X(s), X=A,B,C$
(from top to bottom) found in the incommesurate, $L=25$ system. 
The similarities and differences between the
comensurate and incommensurate cases are discussed below. The lines in
the phase diagram for the incommensurate $L=15$ case are within error bars the
same as those shown in Fig.~\ref{nfig1b}, so that the differences
between Figs.~\ref{nfig1} and \ref{nfig1b} must be attributed
to different commensurability rather than to different sample size.

There are certain trends that can be easily discerned 
\vbox{
\begin{figure}
\includegraphics{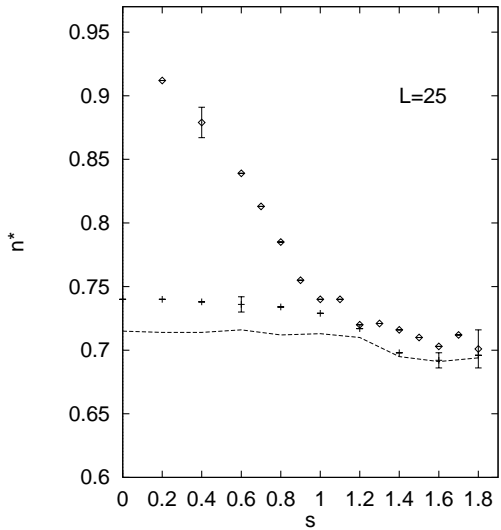}
\vspace{8.5cm}
\caption{The overall phase diagram for the
incommensurate case at size $L=25$ as explained in the text.
The diamonds represent the $A$ line, the crosses the $B$ line and
the dashed line is the $C$ line. Sample error bars have been indicated.
They reflect sample-to-sample variations for six to twelve 
(the number increases with $s$) realizations of the disorder.}
\label{nfig1b}
\end{figure}}
when one follows a free energy minimum as $s$ is increased at constant $n^*$.  If one
starts from the uniform liquid minimum at $s=0$ and a relatively small
value of $n^*$, the free energy value at the minimum (initially zero
according to our convention) decreases steadily with increasing $s$.
For the case in Fig.~\ref{nfig1b} 
at $n^*=0.66$, for example, 
$\beta F$ is close to $-180$ at $s=1.8$.
The density distribution becomes progressively less uniform, with
$\rho_{max}$, which at $s=0$ equals the
average value $\bar{\rho} = \rho_\ell$, rising by more than one order
of magnitude as $s$ increases from zero to one. For a deep glassy state
at a relatively large value of $n^*$, the free energy is strongly
negative even at $s=0$, and its value decreases
further as $s$ increases.  For example, at $n^*=0.78$ the glassy minimum 
for $L=25$ with
$\beta F=-63$ at $s=0$ can be continued  to a minimum with
$\beta F$ equal to $-231$ at $s=1.8$.
The density distribution at a glassy minimum is considerably more
inhomogeneous than that of the liquid minimum continued to the same
value of $s$ and it is less
sensitive to the value of $s$: the quenched disorder has less effect on
a state that is inhomogeneous and disordered to begin with. 

These trends in the behavior of liquid and glassy minima as $s$ is increased
from zero are clearly illustrated by examining the pair correlation
function $g(r)$, defined in
\vbox{
\begin{figure}
\includegraphics{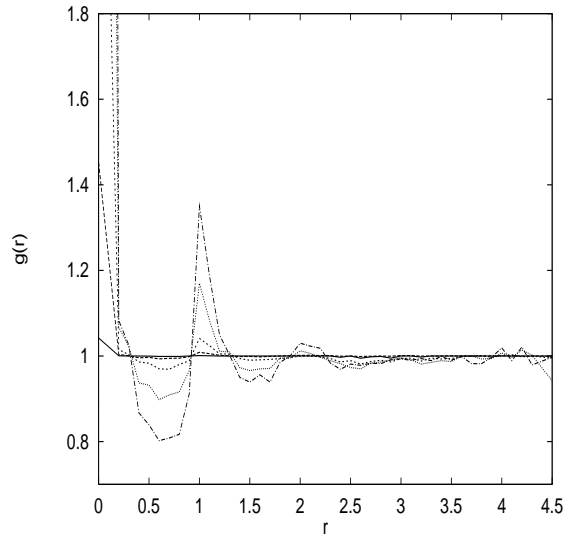}
\vspace{8.5cm}
\caption{Liquid phase correlations. The pair correlation function $g(r)$
as defined in Eq.~(\protect{\ref{gofr}}) plotted as a function of $r$, the
distance in units of $\sigma$, for the liquid-like
minimum at density $n^*=0.66$. The curves shown, in order of
increasing peak height at $r=1$, correspond to $s=0.2, 0.6, 1.0, 1.4, 1.8$. 
The system size is $L=25$.}
\label{nfig2}
\end{figure}}
\vbox{
\begin{figure}
\includegraphics{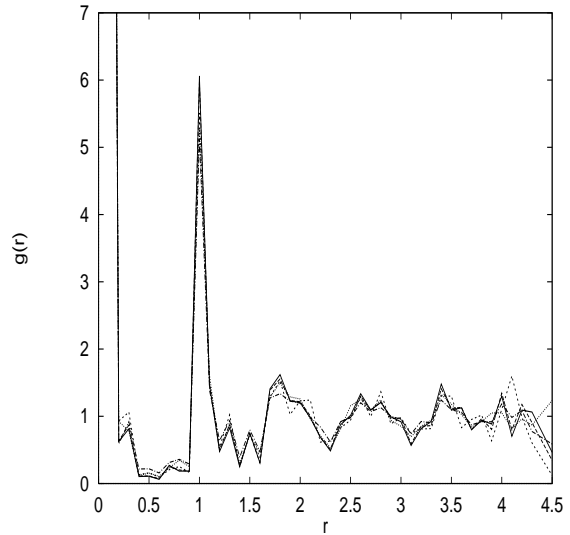}
\vspace{8.5cm}
\caption{The pair correlation function $g(r)$
for a glassy minimum. The curves
shown correspond to the same values of $L$ and $s$ as in 
Fig.\protect{\ref{nfig2}}, but for a glassy minimum at $n^*=0.78$ as
discussed in the text.}
\label{nfig3}
\end{figure}}
Eq.(\ref{gofr}), at each minimum.  In
Fig.~\ref{nfig2}, we show $g(r)$ computed for the liquid minimum at
size $L=25$ and density $n^*=0.66$. The curves shown, in order of
increasing value of the peak near $r=1$, correspond to increasing
values of $s=0.2, 0.6, 1.0, 1.4, 1.8$. There is a clearly visible
increase in structure, which becomes more evident as the value of $s$
increases beyond unity. However, this level of structure is still
quantitatively different from that found for glassy minima at
relatively high densities. This can be seen by comparing
Fig.~\ref{nfig2} with Fig.~\ref{nfig3} where we plot $g(r)$ for a
$L=25$ glassy minimum continued from $s=0$ to $s=1.8$ at $n^*=0.78$,
as mentioned in the preceding paragraph.
We see that the $s$-dependence of the structure is now much weaker, and
the heights of the peaks at $r=0$ and near $r=1$ are much larger than
those in Fig.~\ref{nfig2}. 
These results can be compared to those found in the
replica calculation~\cite{thal}.
To make contact with those results, our $g(r)$ for
the liquid-like minimum should be compared with the function $g_0(r)$
of the replica symmetric solution, and our $g(r)$ for a glassy minimum
with the function $g_1(r)$ of the
replica-symmetry-broken solution. 
Although,  due to differences in the modeling of
the random potential and effects of discretization in the present study
(some of these effects are discussed in section \ref{discuss}), a
detailed, quantitative comparison of our results with those of
Ref.\onlinecite{thal} is not possible, it is clear that the main
features we have discussed are qualitatively similar. 

The crystalline minimum obtained for $s=0$ in commensurate systems at
sufficiently high densities shows very little change in structure as it
is followed to non-zero values of $s$. 
Any effects of weak pinning disorder on the
crystalline order may be too subtle~\cite{larkin,GiaLeD}
to show up at the system sizes and discretization scales used
here in the commensurate case.

\subsection{Instability of the liquid minimum}

We consider first the $A$ line, that is, the density at which the
liquid minimum becomes locally unstable as $n^*$ is increased from a low
initial value, keeping $s$ fixed. This transition is detected at any
desired value of $s$ in the
following way. At a density previously
determined to be well below the value of $n^*_A(s)$ (this determination
is easily performed by trial and error), one ``follows'' the
$s=0$ liquid minimum, as previously explained, to the value of
the disorder strength being studied. The density
configuration at this minimum is the initial condition.
Then, one proceeds to increase
$n^*$ by small intervals, thus moving up along a vertical line in the
phase diagram. At every value of $n^*$ that is reached, we run
our minimization routine (using the configuration at the minimum
obtained at the previous step as the starting point) to locate the
nearest minimum. The density configuration at the minimum is 
analyzed and then used as the initial condition to
study the next higher density. 

In the initial stages of this process, the system remains in the 
liquid-like minimum,
with little change in its properties. However, as $n^*$ reaches the
value $n^*_A(s)$, discontinuities are found.
These are more prominent for the larger
system sizes (Fig.~\ref{nfig1b}) and particularly
dramatic for values of $s$ not too large. 
The free energy drops 
abruptly, as the liquid minimum becomes unstable and the system has to 
find some other nearby minimum (our numerical minimization
procedure is designed to converge only to stable local minima of the
free energy). Computationally, this is heralded by a very
sharp and obvious increase in the number of iterations required by our
numerical procedure to find the free energy minimum nearest to the
starting configuration. This new minimum is invariably glassy, as one
might expect, since
a considerable number of glassy minima are close in configuration space
to the liquid-like minimum~\cite{land}. The value of the free energy at
the minimum 
that the system has reached drops sharply as the $n^*_A(s)$ value
is crossed, because the free energies of glassy minima are considerably
lower in the region of the $(n^*,s)$ plane being considered.
Also, every measure of structure in the system inceases abruptly, since,
as discussed above in connection with Figs.~\ref{nfig2}
and \ref{nfig3},
glassy states are much more inhomogeneous than the liquid-like
ones in this region of the $(n^*,s)$ plane.

An example of the behavior found  is displayed 
in Figs.~\ref{nfig4} and \ref{nfig5}. In the main part of
Fig.~\ref{nfig4}, we show the evolution of the free energy as $n^*$ is
increased
\vbox{
\vspace{0.5cm}
\epsfxsize=8.0cm
\epsfysize=8.0cm
\epsffile{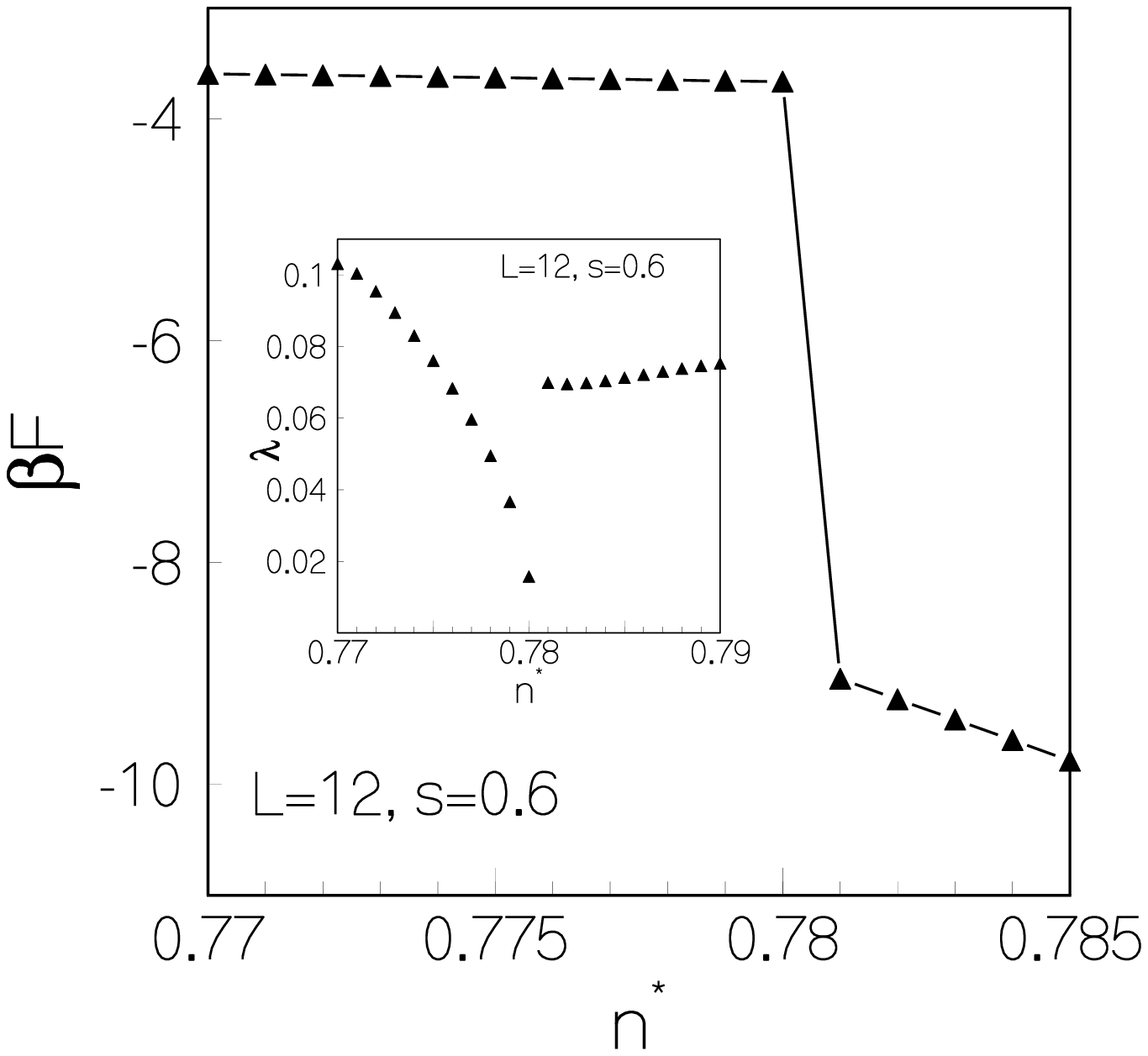}
\begin{figure}
\caption{Discontinuities at the $A$ line. In the main plot, the free energy
in dimensionless form for a $L=12$ sample ($s=0.6$) is plotted 
as a function of density. A sharp drop in the free energy is seen as
the liquid minimum becomes unstable and 
the system switches to a glassy minimum. As shown in the inset, this
switch is also reflected
in the discontinuity in $\lambda$, the smallest eigenvalue of the matrix
${\bf M}^\prime$ defined in Eq.(\protect{\ref{hess2}}).}
\label{nfig4}
\end{figure}}
in steps of 0.001, keeping $s$ fixed at 0.6 for a $L=12$ sample.
One can clearly see that $\beta F$ varies little while the system remains 
in the liquid minimum and jumps abruptly as this minimum becomes
unstable near $n^*_A \simeq 0.78$. 
The behavior for the larger 
incommensurate samples is quite similar, the main difference
being that  the drop in $\beta F$ is much larger, and that the transition
occurs, at this value of $s$, at $n^*_A \simeq 0.84$ for both
$L=15$ and $L=25$. The value of $n^*_A$
can readily be found to very high precision and it
varies little as one averages over different realizations of the
quenched disorder, for the same $s$. The error bars shown in
Fig.~\ref{nfig1b} correspond to an average over six to twelve
realizations (the larger number at larger $s$.)
The results in Fig.~\ref{nfig1} are averages over five realizations.
In the inset, we show that the smallest eigenvalue,
$\lambda$, of the matrix ${\bf M}^\prime$  defined in Eq.(\ref{hess2})
approaches zero as $n^*$ approaches $n^*_A$ from below. This is as
would be expected -- as noted in section \ref{methods}, the instability
of a local minimum is signalled by the vanishing of $\lambda$. 
\vbox{
\vspace{0.5cm}
\epsfxsize=8.0cm
\epsfysize=8.0cm
\epsffile{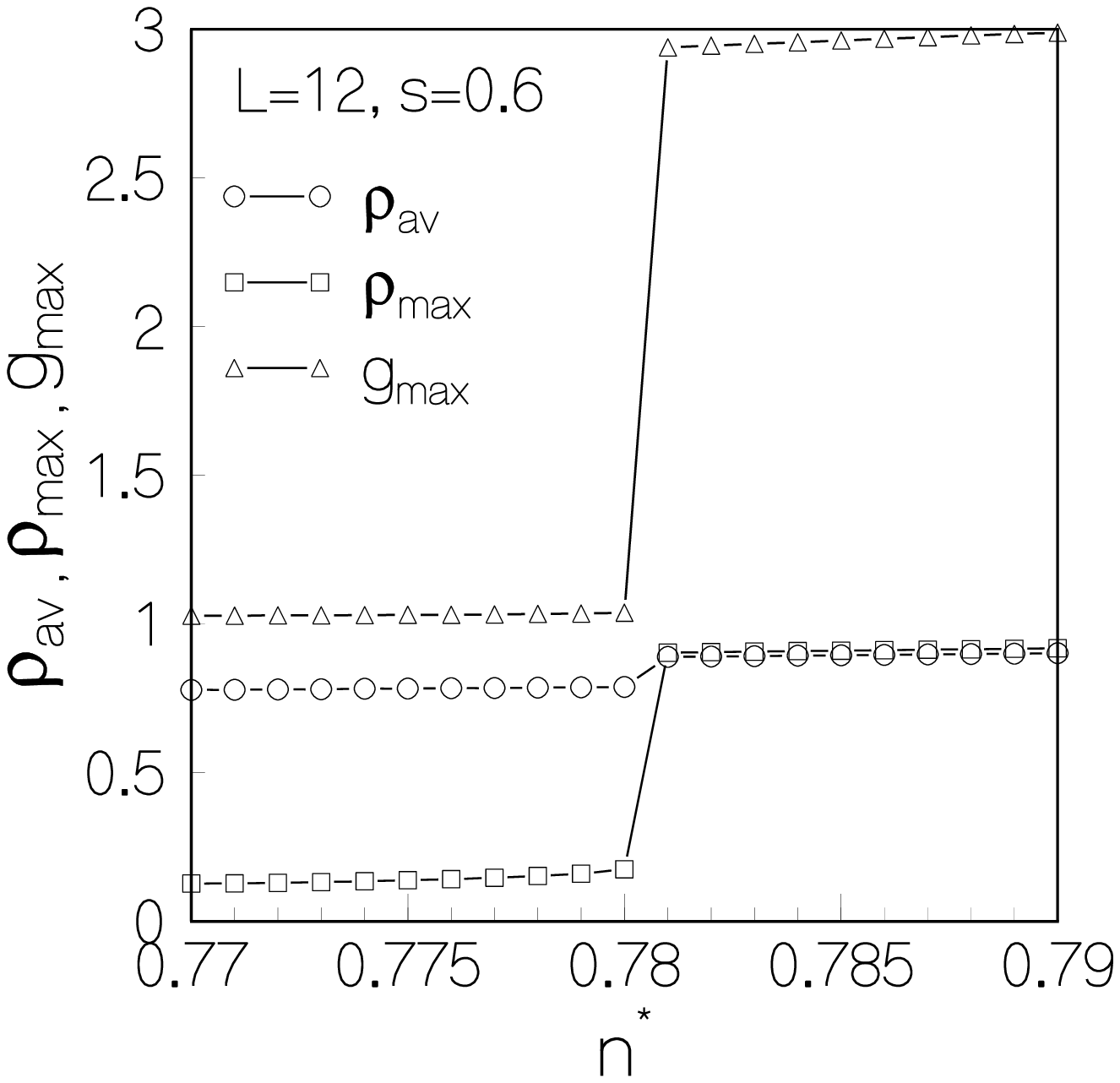}
\begin{figure}
\caption{Example of
how the system becomes more structured as the $A$ line is crossed for an
$L=12$ sample at $s=0.6$. The height $g_{max}$ of the
first finite-$r$ peak in $g(r)$ increases discontinuously, the density 
nonuniformity represented by $\rho_{max}$  exhibits a large
increase, and the average density $\rho_{av}$ shows a small
discontinuous increase. In order to be able
to use a single vertical scale, we have
displayed $\rho_{av}$ rather than $\bar{\rho}$ (see text).}
\label{nfig5}
\end{figure}}
In Fig.~\ref{nfig5}, we show three quantities which characterize the
nature of the density distribution at a minimum. These are: 
$g_{max}$, the value of the pair correlation function $g(r)$
at its first finite $r$ maximum (near $r=1$); 
$\rho_{max}$, the maximum value of the
$\rho_i$; and the dimensionless average density $\rho_{av}$ defined in
section \ref{methods}.
All these quantities exhibit discontinuous changes as the system switches
minima at $n^*=n^*_A \simeq 0.78$, (or at $n^*_A\simeq 0.84$ 
in the incommensurate case).
$g_{max}$ remains close to unity 
as long as the system
stays in the liquid state, and then jumps
to a substantially larger value consistent with the increased
short-range order present in a glassy minimum. 
This can also be seen from Figs.~\ref{nfig2} and \ref{nfig3}.
The value of $\rho_{max}$
also increases by a considerable amount, indicating the increased
inhomogeneity of a glassy minimum relative to the liquid-like one.
The small increase in the value of $\rho_{av}$ reflects that
the average density at a glassy minimum is slightly higher than that at
the liquid-like minimum.

The behavior discussed above changes as $s$ is increased. The change occurs
near $s=1$ for $L=12$, and at somewhat larger $s$ for the other system sizes,
as the $A,B,C$ lines become very close.
The results obtained in the larger-$s$ regime are described in the next
subsection.

\subsection{Instability of glassy minima and the liquid to glass
transition} 

To find the $B$ and $C$ lines, we must start with a carefully chosen
glassy configuration at a relatively high $n^*$ and
fixed $s$, and then follow this configuration 
to lower densities by decreasing $n^*$ in small steps 
($\delta n^* \simeq 0.001$),
keeping the value of $s$ unchanged. This is continued until the minimum
becomes unstable and the minimization routine converges to a new
minimum which, if the starting minimum is chosen
as described below, turns out to be the liquid-like one. 
The density at which this occurs defines the value
of $n^*_C$. Then, comparing
the free energy of the glassy minimum with that of the liquid minimum
obtained for the same realization of the disorder, it is easy to
determine the value of $n^*_B$ -- this is the value of $n^*$ at which
the two free energies are equal.

The determination  of the appropriate starting glassy minimum
is nontrivial. Glassy minima for $s \ne 0$ are obtained 
by continuation from those of the pure system ($s=0$). 
One may think that the best choice would be to take 
the glassy minimum with the lowest free energy at the 
starting ($n^*,s$) point. 
In practice, this is difficult to implement because an
exhaustive enumeration of all the glassy minima is
computationally very hard. The glassy minimum
with the lowest free energy at a particular point in the $(n^*,s)$ 
plane does not in general continue to have the lowest
free energy as the values of $n^*$ and $s$ are changed.
Also, in the pure system, all the configurations
obtained by applying one of the symmetry operations of the
computational mesh to the density
configuration at a particular glassy minimum  also correspond to local 
minima with exactly the same free energy.
For $s \ne 0$, all these symmetry-related minima have to be
considered separately because the presence of the random potential
destroys the symmetries present in the pure limit. 

We have not found a rigorous
solution to this problem. Instead, we first
carried out an exploratory study of how the 
locations of the $B$ and $C$ lines in the phase diagram depend on the choice
of the initial glassy minimum. 
The following choices, among others, were considered in our
initial exploration. 
(a) One of the low-lying $s=0$ glassy minima, continued to finite $s$.
(b) Beginning with the same starting configuration as in (a)
and a specific realization of the random
variables $\{V_i\}$, minimize the random potential energy (the
last term in Eq.(\ref{discr})) with respect to all symmetry operations
of the computational mesh. This  attempts to find the
configuration that minimizes, among
all the symmetry related ones, the contribution
of the random potential to the free energy but it is not quite
rigorous because the minimization is
performed  using the values of $\{\rho_i\}$ {\em at the
$s=0$ minimum}.
(c) The glassy minima to which the system 
moves when the density is increased above the $A$ line, as discussed in
the preceding subsection. 

The outcome of this study is that the
locations of the $B$ and $C$ lines in the $(n^*,s)$ plane are not 
sensitive to the choice of the glassy minimum as long as it is one of
the low-lying minima. (Even when we have deliberately 
or accidentally chosen a ``wrong'', non-low-lying, minimum, we have
found that the system
often spontaneously makes a glass-to-glass transition~\cite{comment2}
to a low-lying
minimum as one decreases $n^*$ above the $B$ line.)
The variation of the values of 
$n^*_B$ and $n^*_C$ for different choices of the glassy minimum is
comparable to the uncertainty of these values arising from
sample-to-sample variations  caused by differences in the realization
of the disorder. The results described below were
obtained (unless otherwise indicated) from runs in which a low-lying
glassy minimum obtained from
continuation of one at $s=0$ was taken to be the initial state
for the density-lowering run.
As explained at the beginning of
this subsection, the initial configuration is
followed to lower densities at fixed $s$ and the $n^*_C(s)$
and $n^*_B(s)$ points are found for that value of $s$.
For relatively small values of $s$, the signatures of the $C$
instability are very easy to detect: they are similar to the
discontinuities shown in Figs.~\ref{nfig4} and \ref{nfig5}. At larger
values of $s$ more care is required.

For small values of $s$, the $A$, $B$ and $C$ lines are well separated from
one another. However, as the value of $s$ is increased, these three
lines begin to approach each other. As shown in Figs.~\ref{nfig1}
and \ref{nfig1b}, the
separation between lines $A$ and $B$ decreases rather rapidly with
increasing $s$, while the separation between lines $B$ and $C$ decreases
more slowly. Finally, near $s=1$, these three lines appear to merge
with one another for the $L=12$ system. For $L=25$ (and also for $L=15$),
the separation between them does not exceed the combined error bars,
but separate $B$ and $C$ transitions can be detected
in most (not all) ``runs'' (i.e. realizations
of the disorder) as explained in detail below. At
larger $s$, it becomes increasingly difficult to resolve these
three lines as they come close to each other.
Since lines $A$ and $C$ represent, respectively, the limits of stability
of the liquid and glassy minima and line $B$ corresponds
to the first-order liquid-glass transition, a merging of these three
lines suggests that this transition becomes continuous as $s$ is
increased beyond a ``tricritical'' value which would be close to unity for
the $L=12$ commensurate sample and somewhat larger for the incommensurate
samples. Another possibility is that the first-order
liquid-glass transition disappears beyond a critical point near
$s=1$. 

To examine the behavior in this region more closely,
we have carried out several numerical experiments in
which the value of $n^*$ is ``cycled'' through the liquid-glass
transition, keeping $s$ fixed at values close to unity. 
In this way, the three lines are detected in the same ``run''.
These numerical
experiments are similar to simulations of hysteresis in magnetic phase
transitions. We start with the liquid minimum at a low value of $n^*$ 
(below line $C$), and increase $n^*$ in small steps, keeping $s$ fixed.
The liquid minimum is thus followed to higher densities until it
undergoes a rapid change signalling a possible instability. The process
of increasing $n^*$ in small increments is continued for a few more
steps, and then the local minimum so obtained is followed to lower
densities by decreasing $n^*$ is small steps. This is continued until
the starting value of $n^*$ is reached. If the liquid-glass transition
at the chosen value of $s$ is first-order with the three densities
$n^*_A$, $n^*_B$, and $n^*_C$ separated from one another, then the
cycling experiment described above should exhibit clear evidence of
hysteresis. This is indeed what we find, for
\vbox{
\vspace{0.5cm}
\epsfxsize=8.0cm
\epsfysize=8.0cm
\epsffile{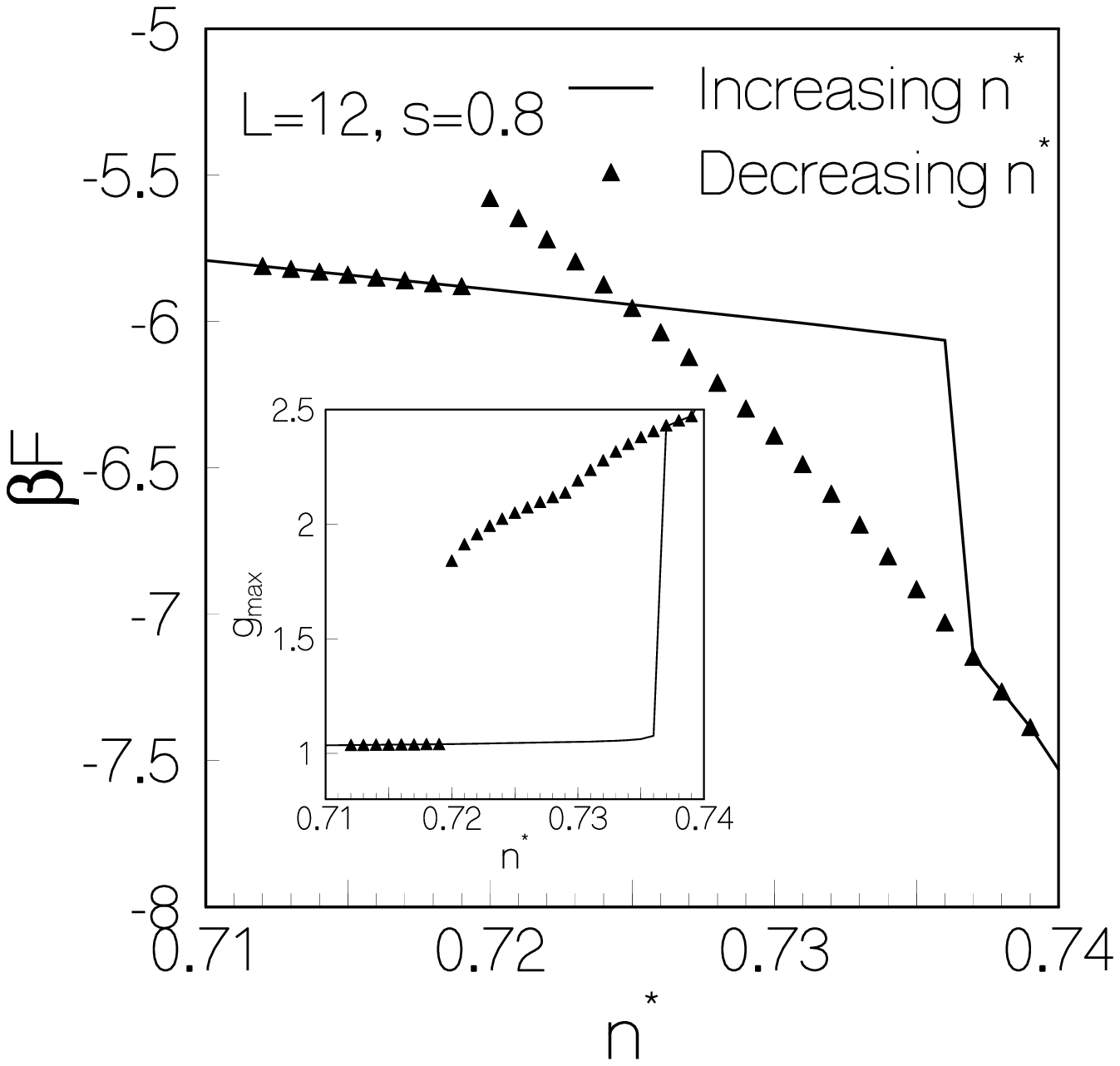}
\begin{figure}
\caption{Hysteresis and discontinuities across the liquid-glass transition
at small values of $s$. In the main plot, the dimensionless free energy
of the stable minimum is plotted vs. density as one cycles across the
$A$, $B$ and $C$ lines as explained in the text. Hysteresis is clearly
observed. In the inset, the quantity $g_{max}$ is shown. The results
shown are at $s=0.8$ for a commensurate $L=12$ system, but the same
behavior is found in this range of $s$ for incommensurate systems.}
\label{nfig6a}
\end{figure}}
all system sizes and at every run, if the value of $s$ is lower
than a certain critical value. A typical example is shown in
Fig.~\ref{nfig6a} which shows the results for a $L=12$ sample at
$s=0.8$. The hysteresis in the free energy and $g_{max}$ (shown in
the inset) is evident: the liquid minimum becomes unstable at $n^*_A
\simeq 0.735$ as $n^*$ is increased from a low initial value, while the
glassy minimum found for $n^* > n^*_A$ can be continued all the way
down to $n^*_C \simeq 0.720$ before it becomes unstable. The
liquid-glass transition occurs at $n^*_B \simeq 0.725$ where the two
branches of the free energy cross. The same situation occurs for
the incommensurate  $L=25$ system except that the values  of 
the transition points are $n^*_A \simeq 0.79$, $n^*_B \simeq 0.73$
and $n^*_C \simeq 0.71$ for $s=0.8$. The results at $L=15$ are, within
error bars, the same as those for $L=25$ at this value of $s$.
 
The behavior in Fig.~\ref{nfig6a} is to be contrasted with that
shown in Fig.~\ref{nfig6b} which displays the results of the cycling
experiment on a $L=12$ sample at $s=1$. The distribution of
the random variables $\{r_i\}$ in this sample is the same as that of
Fig~\ref{nfig6a} -- only the strength of the disorder is changed.
In this figure, there is no evidence of
\vbox{
\vspace{0.5cm}
\epsfxsize=8.0cm
\epsfysize=8.0cm
\epsffile{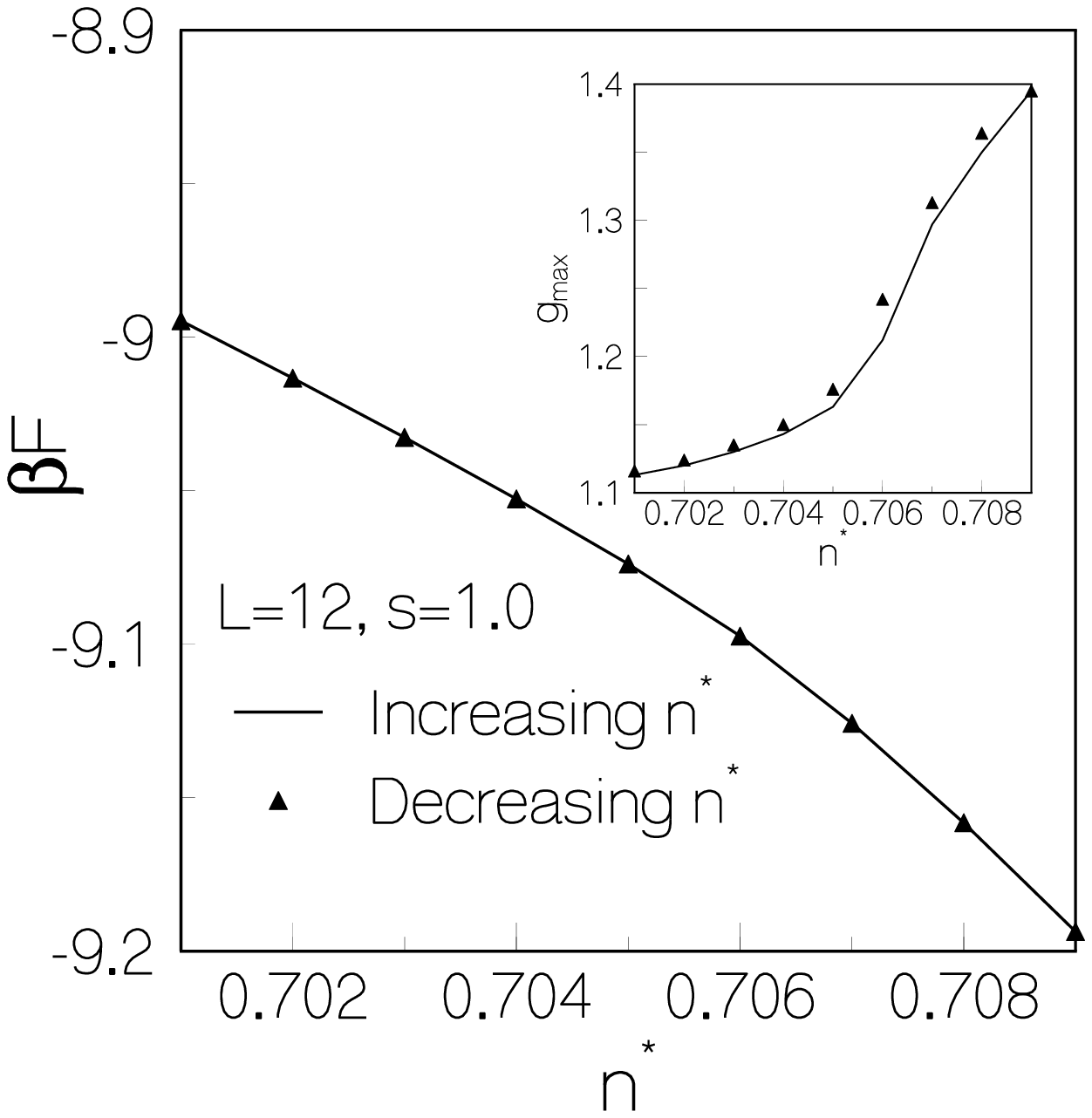}
\begin{figure}
\caption{Cycling across the liquid-glass transitions for $s$ = 1
in a $L=12$ commensuarte system. The same quantities are plotted 
as in Fig.~\protect{\ref{nfig6a}}, and now no hysteresis is seen. 
Incommensurate systems exhibit the same behavior at somewhat larger
values of $s$, but not in all runs.}
\label{nfig6b}
\end{figure}}
hysteresis in the
free energy. The plot of $g_{max}$ shown in the inset exhibits a sharp
change near $n^* = 0.706$ for both inreasing-$n^*$ and decreasing-$n^*$
runs, and the results for the two runs are nearly identical. Given the 
rounding off errors associated with 
the numerical procedures we use,
the small differences between the increasing-$n^*$ and decreasing-$n^*$
values of $g_{max}$ are likely to be insignificant. We, therefore,
conclude that at least within the resolution of our numerical
procedures, there is no hysteresis at $s=1.0$ for this $L=12$ sample. 
This implies that the first order transition found in this sample for
$s=0.8$ either becomes a continuous one or disappears as the value of 
$s$ is increased to 1.0. The sharp change in the value of $g_{max}$ 
near $n^*=0.706$ suggests that the transition persists as a continuous
one. To investigate this further, we have calculated the derivatives of
$g_{max}$, $\rho_{max}$, and $\rho_{tot} \equiv \sum_i \rho_i$ with
respect to $n^*$ in the region where these quantities change rapidly.
We have also examined the behavior of $\lambda$
as a function of $n^*$ in this region. 
\vbox{
\vspace{0.5cm}
\epsfxsize=8.0cm
\epsfysize=8.0cm
\epsffile{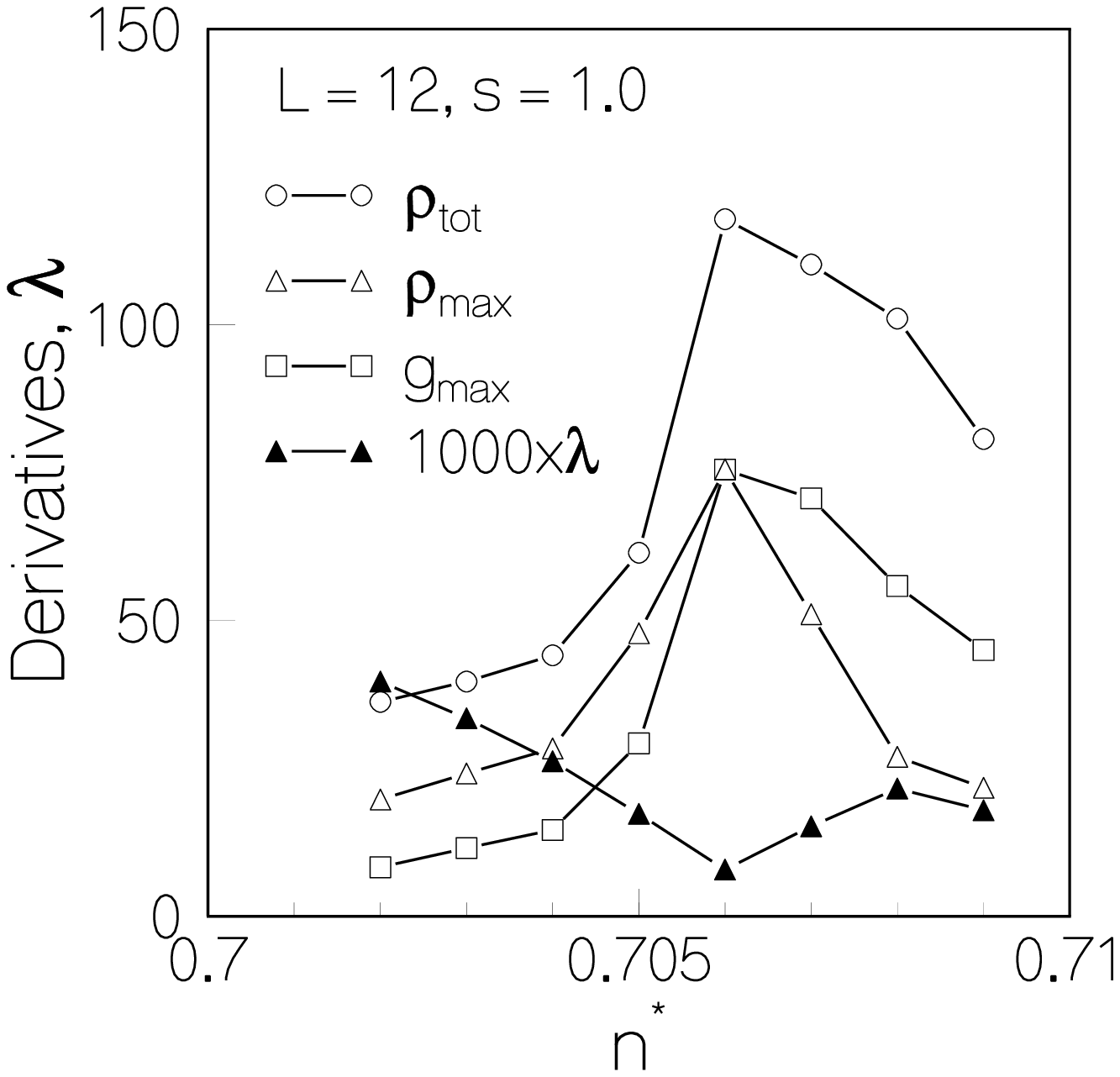}
\begin{figure}
\caption{Derivatives with respect to $n^*$ of
the quantities $\rho_{tot}$, $\rho_{max}$ and
$g_{max}$, as defined in the text, plotted as functions of
$n^*$ across a putatively continuous liquid-glass
transition in a $L=12$ sample with $s=1.0$. The three quantities have 
sharp peaks at $n^*=0.706$.
The eigenvalue $\lambda$, also defined in the text,
shows a pronounced dip at the same point.}
\label{nfig6c}
\end{figure}}
Results for these quantities are shown in Fig.~\ref{nfig6c} for
the same sample as that of Fig.~\ref{nfig6b}. 
All the derivatives exhibit sharp peaks at $n^*=0.706$, and the value
of $\lambda$ goes through a minimum that is very close to zero at the
same point. These results strongly suggest the occurrence of a
continuous phase transition at $n^*=0.706$. However, due to the limited
resolution of our numerical calculations and the smallness of sample
size, we can not rule out the possibility that the observed behavior
reflects a sharp crossover rather than a true phase transition. Similar
results are found for larger values of $s$. The continuation of the
``transition line'' beyond the point where the lines $A$, $B$ and $C$ come
together is determined by locating the value of $n^*$ at which the
eigenvalue $\lambda$ reaches a minimum.
The value of $s$ at which the $A$, $B$ and $C$ lines merge and
the hysteresis in the cycling experiment disappears is found to be
weakly dependent on the realization of the disorder -- it varies
between 1.0 and 1.2 for the five different $L=12$ samples studied.

For the incommensurate samples, the situation is
somewhat more ambiguous. For $L=25$, the same
cycling procedure shows that the transition is clearly hysteretic
for all runs with $s\le 1.1$. For larger values of $s$, an increasingly
larger percentage of the runs is non-hysteretic (i.e. the results
for $\beta F$ look like those in Fig.~\ref{nfig6b}), while the other
runs display a behavior similar to that in Fig.~\ref{nfig6a} but
with much smaller discontinuities. As $s$ is increased beyond $s=1.8$, it
becomes, in most of the ``runs'', impossible to distinguish the
discontinuities, if any, from computer noise. Thus, it is possible
in this case to plot separate $A$, $B$, and $C$ lines all the way
up to $s=1.8$. This accounts for the obvious difference 
in this respect between Figs.~\ref{nfig1} and \ref{nfig1b}. 
The results for $L=15$ are quite consistent with those for $L=25$, 
but the smaller system size makes all interpretations more difficult. 
Thus, it is more difficult to identify 
the precise position of any well-defined tricritial point (or a
critical point) from the  results for the incommensurate samples.
One might alternatively say that these incommensurate results 
are indicative of a crossover.
It is not
possible to completely rule
out that the behavior is different for the commensurate and
incommensurate samples, or that the 
poorer resolution of the smaller samples masks discontinuous behavior
in some of the larger $s$ runs.

\subsection{Crystallization}

To study how the crystallization density $n^*_D$ changes as $s$ is
increased from zero (i.e. the location of the line $D$), we start with
the crystalline minimum obtained for a commensurate sample at
a large value 
\vbox{
\vspace{0.5cm}
\epsfxsize=7.5cm
\epsfysize=7.5cm
\epsffile{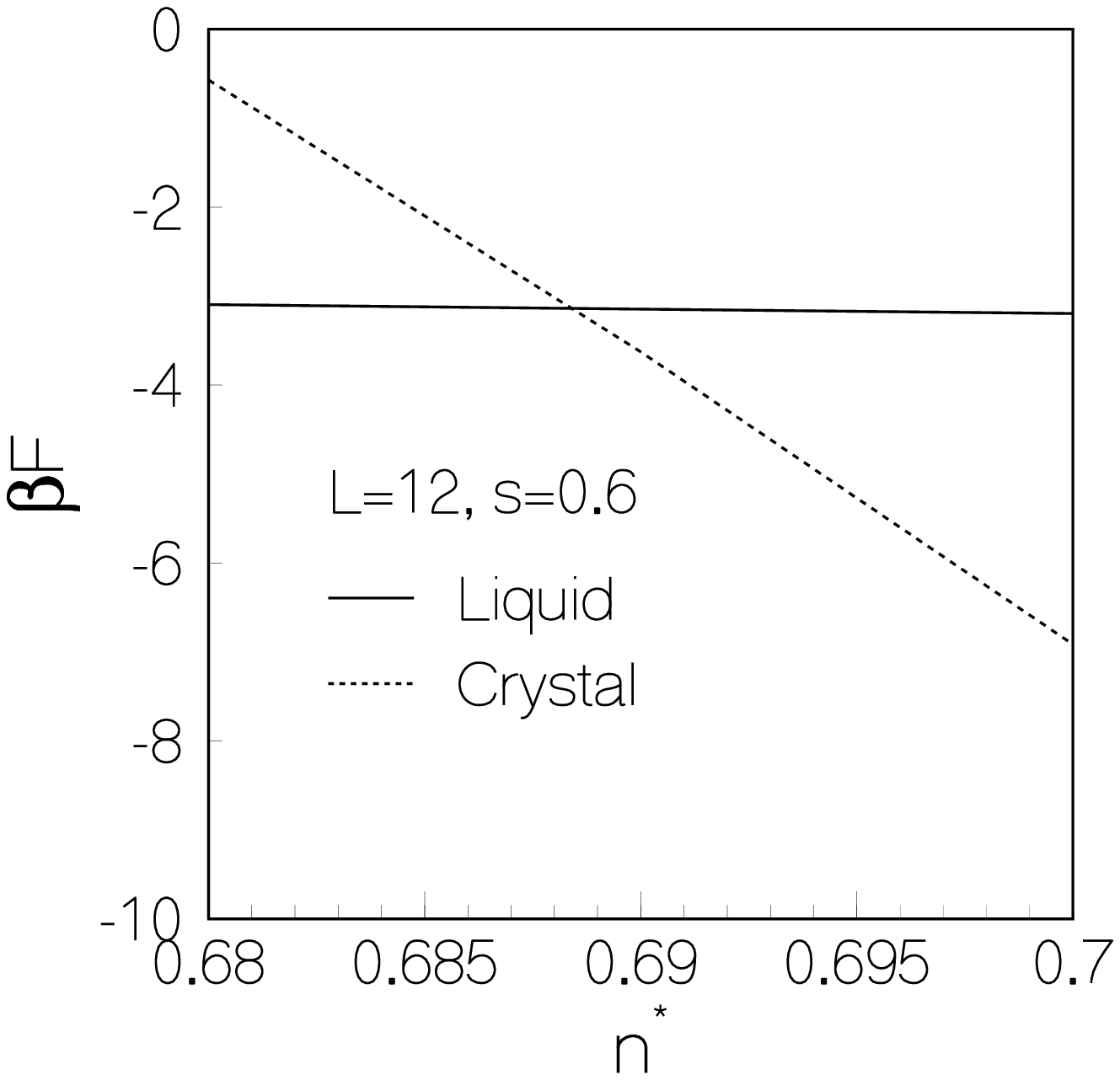}
\begin{figure}
\caption{Free energy crossing at the crystallization transition.
The solid and dotted lines represent, respectively, the free energies
of the crystalline and liquid-like minima of a $L=12$ sample with
$s=0.6$. Their crossing point is the density $n^*_D(s)$.}
\label{nfig8}
\end{figure}}
of $n^*$. We then find the symmetry related configuration
that minimizes the random
potential energy for a particular realization of the
disorder and continue this configuration to the desired value of $s$.
This configuration is then continued to smaller values of $n^*$ by
decreasing $n^*$ in small steps. The crystalline minimum turns out to
be quite robust under changes of the density and the strength of the
disorder -- the minimization routine converges rather quickly to the
new minimum as the value of $n^*$ or $s$ is changed by a small amount.
While decreasing the value of $n^*$, we keep track of the free energy
of the crystalline minimum and find the value of $n^*$ at which this
free energy crosses that of the liquid minimum for the same realization
of the disorder. For relatively small values
of $s$, the crossing point determines
the value of $n^*_D$ for the chosen value of $s$.
Typical results for the crossing of these two free
energies are shown in Fig.~\ref{nfig8}. 
Our results for line $D$, averaged over five
realizations of the disorder, are shown in Fig.~\ref{nfig1}. 
The crystallization transition is strongly first order for all values 
of $s$. In the small $s$ regime, the crystalline minimum has the 
lowest free energy for all densities above line $D$. Therefore, the lines
$A$, $B$ and $C$ do not have any 
equilibrium thermodynamic significance in this regime for a commensurate
system:
the liquid-glass transition at line $B$ can be observed only if the
crystallization transition at line $D$ is  avoided, e.g. by rapid
compression. 

As shown in Fig.~\ref{nfig1}, the crystallization line crosses the
liquid-glass transition line at a point which is very close to that
where the lines $A$, $B$ and $C$ seem to
come together. Beyond this 
point, line $D$ is determined by the crossing of the free energies of
glassy and crystalline minima. The procedure is quite analogous to
that shown in Fig.~\ref{nfig8}.
This line, therefore, represents a first
order transition between crystalline and glassy states in this regime. 
The phase diagram of Fig.~\ref{nfig1} implies that
the system undergoes a first order liquid-to-crystal transition for
small values of $s$ as the density is increased from a low initial
value. However, as the value of $s$ is increased above a critical value
(which is close to unity for the $L=12$ system), the transition as
$n^*$ is increased becomes a continuous liquid-to-glass transition (or
perhaps a sharp crossover). The glassy state then undergoes a first
order transition to the crystalline state as the density is increased
further. The observed curvature of line $D$ for large $s$ also implies 
that the system would undergo a first order crystal-to-glass transition
as the strength of the disorder is increased at constant density.

\section{Summary and Discussion}
\label{discuss}

We have mapped out the mean-field phase diagram of a
hard sphere system in the presence of a quenched random
potential by numerically studying the evolution of the minima of
a model free energy as a function of
the density $n^*$ and the strength $s$
of the disorder. The phase diagram in the $(n^*,s)$
plane exhibits liquid, glassy and crystalline (for commensurate
samples) phases. We find that the standard first order crystallization
transition which occurs  at $s=0$ upon increasing $n^*$
retains its character at small $s$ as a first order transition
from a weakly inhomogeneous liquid phase to a nearly
crystalline state. The density at
which this transition occurs increases somewhat with the strength of the
disorder. We also find for all samples 
a liquid to glass transition 
in the metastable, ``supercompressed'' regime.
This transition is first-order for small $s$, but within
the resolution of our results it appears to become
continuous as $s$ is increased above a critical value. This
critical value is larger for incommensurate samples. The
crystallization line in the $(n^*,s)$ plane crosses the glass
transition line near the point where the glass transition becomes
continuous. Thus, the first order crystallization transition
found for small $s$ as the density is increased from a small
initial value is replaced, for sufficiently large $s$, by a
continuous liquid to glass transition. The phase diagram also
shows, at larger $s$, a first order crystal to glass transition as the
strength of the disorder is increased at constant density.

All the qualitative features of our phase diagram (i.e. its
topology, the shapes of the transition and instability lines,
and the nature of the transitions) are identical to those found
in the replica-based analytic study~\cite{thal} of the same
syatem~\cite{comp}. Two of our most important
results, namely the change in the nature of the liquid to glass
transition as the strength of the disorder is increased above a
critical value, and the crossing of the crystallization and
glass transition lines above this critical value of the disorder
strength, were also found in the analytic calculation. This
similarity between the results of two 
studies using extremely different
methodologies strongly suggests that the qualitative features of
our phase diagram are correct, at least at the mean-field level.
The considerable
quantitative differences that exist between the numerical
and replica  results, i.e. that all
the transition and instability lines in our phase diagrams lie
at substantially lower densities than those 
obtained in the analytic study, have the same origin as the discrepancy
between our $s=0$ results and those of molecular dynamics
simulations~\cite{md} of the pure hard sphere system. As noted
in our earlier work~\cite{cdotv96,sit}, these differences result
from  the discretization of the free energy
functional. The use of a simple cubic mesh of spacing $h \sim 0.2
\sigma$ in the discretization procedure increases the relative
stability of inhomogeneous local minima of the free energy and
thus leads to substantially lower values for the densities at
which crystallization and the glass transition occur.
The quantitative differences between our results in
Fig.~\ref{nfig1} and those in
Fig.~\ref{nfig1b}, on the other hand, appear to arise chiefly from the 
incommensurability  of the latter sample, rather than from the slight
difference in the values of $h$, or even from that in
the values of $L$: we have found negligible sample-size
effects in comparing the $L=15$ results to those at $L=25$
at the same value of $h$. 
The effects of discretization would presumably disappear for $h$
much smaller than the width of the approximately Gaussian
density distributions near the points where the particles are
localized at an inhomogeneous minimum of the continuum free
energy functional. Unfortunately, a numerical calculation with
such small values ($\sim 0.01 \sigma$) of $h$ would require
dealing with a very large number (of the order of $10^6$) of
variables $\{\rho_i\}$. This appears to be 
computationally intractable. 

Our phase diagram is a mean-field one
-- possible effects of fluctuations are not included in our
calculation. The first-order crystallization transition is not
expected to be strongly affected by fluctuations. The situation
is more complex for the glass transition because there are a
large number of glassy local minima. When the effects of
fluctuations are included, the system might visit a large number
of different glassy minima during its evolution over a long
period of time, and thus behave like a liquid in the sense that
the particles would no longer be localized in space and the
time-averaged local density would be only weakly inhomogeneous. A true
thermodynamic glass transition would occur only if the
characteristic time scale for transitions between different
glassy minima diverges in the thermodynamic limit. Whether
this happens in the pure system is still a highly controversial
issue. Further investigations of this question
for systems of particles in the presence of quenched disorder
would be very worthwhile. Also, the presence of multiple
low-lying glassy minima of the free energy is expected to lead
to slow relaxation even if no thermodynamic glass transition is
present. Therefore, signatures of the mean-field glass
transition found in our study should show up in the dynamics of
the system even if no thermodynamic glass transition occurs when
fluctuations are taken into consideration.

Our density--disorder phase diagram exhibits
qualitative similarities with the 
field--temperature phase diagram of 
some high-$T_c$ superconductors in
the presence of random point pinning. 
For a system of vortices in the mixed phase of
type-II superconductors, the temperature $T$ plays the role of
the density $n^*$ of the hard sphere system -- increasing $T$ is
analogous to decreasing $n^*$. As pointed out in the
Introduction, increasing the
magnetic field $H$ is believed~\cite{Safar} to increase the
effective strength of the pinning disorder. Using these
analogies, one can translate, in a very crude and
qualitative sense, our phase diagram in the $(n^*,s)$ plane to a
phase diagram for superconductors in the $(T,H)$ plane. Then, our
result that the crystallization transition at weak disorder is
replaced by a continuous glass transition as the strength of the
disorder is increased translates into the
statement  that for superconductors, the
first-order liquid to Bragg glass transition at low fields
should change over to a continuous glass transition as the field
is increased. As noted in the Introduction, this is precisely
the behavior found in experiments on 
a family of high-$T_c$ superconductors.
The experimentally obtained phase diagram of these
superconductors also exhibits a Bragg glass to amorphous solid
transition as the field $H$ is increased at low temperatures.
This is analogous to the crystal to glass transition found in
our phase diagram as the strength of the disorder is increased at
constant density. Further evidence in support of this analogy is
provided by a recent numerical study~\cite{condmat} that
suggests that the high-field, low-temperature phase of 
high-$T_c$
superconductors (the so-called vortex glass phase) is very
similar to a structural glass. In view of these similarities, an
extension of our calculation to a system of pancake vortices in
layered superconductors with random point pinning, using the
appropriate form of the free energy, would be of obvious
interest.

We are not aware of any experimentally studied
system that provides a direct and
precise physical realization of the model
studied here. Colloidal suspensions in the presence of a
time-independent, spatially random external potential (produced,
for example, by suitably configured laser fields~\cite{laser}) would 
probably provide a close approximation to our model. Since simple liquids
with short range pair potentials which are strongly repulsive at
short distances behave in many ways like a hard sphere liquid,
our calculation is expected to apply, at least qualitatively,
to such systems also. 

\acknowledgments

We are grateful to F. Thalmann, G. I. Menon, A. K. Sood,
S. Bhattacharya, G.F. Mazenko and T. Witten for helpful discussions
or comments.


\begin{references}
\bibitem[*]{chandan} Also at Condensed Matter Theory Unit, Jawaharlal
Nehru Center for Advanced Scientific Research, Bangalore 560064, India.\\
Electronic address: cdgupta@physics.iisc.ernet.in
\bibitem[+]{oriol} Electronic address: otvalls@tc.umn.edu
\bibitem{rev} For a review, see T. Giamarchi and P. Le Doussal in {\it
Spin Glasses and Random Fields}, ed. A. P. Young (World Scientific,
Singapore, 1998).
\bibitem{htsc} G. Blatter {\it et al.}, Rev. Mod. Phys. {\bf 66} 1125, 
(1994).
\bibitem{pormed} E. Lomba, J.A. Given, G. Stell,
J.J. Weis and D. Levesque, Phys. Rev. E {\bf 48}, 233 (1993).
\bibitem{magbub} R. Seshadri and R. M. Westervelt, Phys. Rev. B {\bf
46}, 5142 (1992); {\it ibid.} {\bf 46}, 5250 (1992).
\bibitem{wxtal} E. Y. Andrei, G. Deville, D.C. Glattli, F.I.B. Williams,
 E. Paris and B. Etienne, Phys. Rev. Lett. {\bf 60},
2765 (1988).
\bibitem{larkin} A. I. Larkin, Zh. Eksp. Teor. Fiz. {\bf 58}, 1466
(1970) [Sov. Phys. JETP {\bf 31}, 784 (1970)]; A. I. Larkin and Y. N.
Ovchinnikov, J. Low Temp. Phys. {\bf 34}, 409 (1979).
\bibitem{Nat} T. Nattermann, Phys. Rev. Lett. {\bf 64}, 2454 (1990).
\bibitem{GiaLeD} T. Giamarchi and P. Le Doussal, Phys. Rev. Lett. {\bf
72}, 1530 (1994): Phys. Rev. B {\bf 52}, 1242 (1995).
\bibitem {caveat} In spite of extensive experimental and theoretical 
investigation over several decades, the question of whether a true
thermodymic glass transition occurs in supercooled liquids in the
absence of disorder remains controversial.  In the absence of a clear
answer to this question, one may operationally define a ``glass
transition'' to occur at the temperature below which the characteristic
relaxation time in the liquid exceeds a given value $\tau_c$.
\bibitem{Zeldov} E. Zeldov {\it et al.}, Nature {\bf 375}, 373 (1995).
\bibitem{Safar} H. Safar, P. L. Gammel, D. A. Huse, D. J. Bishop,
W. C. Lee, J. Giapintzakis, and D. M. Ginsberg, Phys. Rev. Lett. {\bf 70}, 
3800 (1993).
\bibitem{F2-Huse} D. S. Fisher, M. P. A. Fisher and D. A. Huse, 
Phys. Rev. B {\bf 43}, 130 (1990).
\bibitem{Koch} R. H. Koch, V. Foglietti, 
W. J. Gallagher, G. Koren, A. Gupta, and M.P.A. Fisher, Phys. Rev. 
Lett. {\bf 63}, 
1511 (1989).
\bibitem{Cubitt} R. Cubitt {\it et al.}, Nature {\bf 365} 407 (1993).
\bibitem{Khaykovitch} B. Khaykovitch, M. Konczykowski, E. Zeldov, 
R.A. Doyle, D. Majer, P.H. Kes, and T.W. Li , Phys. Rev. B {\bf 56}, 
R517 (1997).
\bibitem{comment} The question of whether a true thermodynamic glass
phase exists in superconductors with random point pinning is
controversial - see, e.g. H. S. Bokil and A. P. Young, Phys. Rev.
Lett.{\bf 74}, 3021 (1995).
\bibitem{thal} F. Thalmann, C. Dasgupta, and D. Feinberg, Europhys.
Lett. (in press) [cond-mat/0001424]
\bibitem{MenDas} G. I. Menon and C. Dasgupta, Phys. Rev. Lett.
{\bf 73}, 1023 (1994).
\bibitem{MezPar} M. M{\'e}zard  and G. Parisi, J. Phys. A {\bf 29},
6515 (1996).
\bibitem{book} M. M{\'e}zard, G. Parisi and M. A. Virasoro,
{\it Spin glass theory and beyond.} World Scientific, 1987.
\bibitem{hm86} J. P. Hansen and I. R. McDonald, {\it Theory of Simple 
Liquids} (Academic, London, 1986).
\bibitem{ry} T. V. Ramakrishnan and M. Yussouff, Phys. Rev. B {\bf 19},
2775 (1979).
\bibitem{ErtNel} D. Erta\c{s} and D. R. Nelson, Physica C {\bf 271}, 79
(1996).
\bibitem{cd92} C. Dasgupta, Europhys. Lett. {\bf 20}, 131, (1992).
\bibitem{cdotv96} C. Dasgupta and O.T. Valls, Phys. Rev. E {\bf 53}, 
2603 (1996).
\bibitem{cdotv98} C. Dasgupta and O.T. Valls,  Phys. 
Rev. {\bf E 58}, 801 (1998).
\bibitem{land} C. Dasgupta and O.T. Valls, Phys. Rev. E {\bf 59}, 3123
(1999).
\bibitem{sit} C. Dasgupta and O.T. Valls, in {\it Complex behavior
of Glassy Systems}, ed. by M Rub\'{\i} and C. P\'{e}rez-Vicente,
Springer, Berlin (1997).
\bibitem{nosmall} For the case of the liquid minimum, it is not 
in fact necessary to make the step $\delta s$ very small. One gets 
just the same configuration by directly increasing $s$ from zero to
the desired value, basically dropping the system into the right spot
in the $(n^*,s)$ plane. To follow the glassy minima, however, more care
must be taken. The configurations obtained
are independent of  $\delta s$ if this quantity is below 0.05.
\bibitem{comment2}  We do not attach any significance to the density
at which this glass to glass transition
occurs, and continue to decrease the density until the
new glassy minimum becomes unstable and the system converges to the
liquid minimum.
\bibitem{comp} Our line $A$ should be identified
with the line (labelled IN in Fig.1a of Ref.\protect\onlinecite{thal})
at which the replica symmetric solution becomes unstable, line $B$
with the line labelled TGT in Fig.1a of Ref.\protect\onlinecite{thal}, and
line $C$ with the ``dynamical transition'' line
of Ref.\protect\onlinecite{thal} (labelled DT in Fig.1a there) which
corresponds to the first appearance of the
replica-symmetry-broken solution. The crystallization line has
the same meaning in both calculations.  
\bibitem{md} L. V. Woodcock, Ann. N. Y. Acad. Sci. {\bf 371}, 274 (1981).
\bibitem{laser} Q.-H. Wei {\it et. al.}, Phys. Rev. Lett. {\bf 81},
2606 (1998).
\bibitem{condmat} C. Reichhardt, A. von Otterlo and G. T. Zimanyi, to
be published [cond-mat/9910314].
\end{references}
\end{document}